\documentclass[12pt]{article}
\usepackage{epsfig}
\usepackage{amssymb}

\newcommand{\be}{\begin{equation}}
\newcommand{\ee}{\end{equation}}
\newcommand{\bea}{\begin{eqnarray}}
\newcommand{\eea}{\end{eqnarray}}
\newcommand{\ov}{\overline}
\newcommand{\ve}{\varepsilon}
\newcommand{\eps}{\epsilon}
\newcommand{\ba}{\begin{array}}
\newcommand{\ea}{\end{array}}

\begin{document}

\vspace{10mm}
\centerline{\LARGE\bf Gluon condensate in charmonium}
\vspace{3mm}
\centerline{\LARGE\bf sum rules with 3-loop corrections}
\vspace{10mm}
\centerline{\large B.L. Ioffe and K.N. Zyablyuk}
\vspace{3mm}
\centerline{\tt ioffe@vitep1.itep.ru, zyablyuk@heron.itep.ru}
\vspace{5mm}
\centerline{\it Institute of Theoretical and Experimental Physics,}
\centerline{\it B.Cheremushkinskaya 25, Moscow 117218, Russia}

\begin{abstract}
Charmonium sum rules are analyzed with the primary goal to obtain the restrictions
on the value of dimension 4 gluon condensate. The moments $M_n(Q^2)$ of the polarization operator
of vector charm currents are calculated and compared with experimental data.
The 3-loop ($\alpha_s^2$) perturbative corrections,
the gluon condensate contribution with $\alpha_s$ corrections and dimension 6 operator $G^3$ contribution
are accounted. It is shown that the sum rules for the moments do not work at $Q^2=0$, where the perturbation
series diverges and $G^3$ contribution is large. The domain in the plane $(n, Q^2)$, where the sum rules
 are legitimate, is found. Strong correlation of the values of gluon condensate
and $\ov{\rm MS}$ charm quark mass quark is determined. The absolute limits
are found to be for gluon condensate
$\left<{\alpha_s\over\pi} G^2 \right>=0.009\pm 0.007 \, {\rm GeV}^4$
and for charm quark mass ${\bar m}({\bar m})=1.275\pm 0.015 \, {\rm GeV}$ in $\ov{\rm MS}$ scheme.
\end{abstract}

\centerline{PACS: 14.65.Dw, 11.55.Hx, 12.38.Bx}

%%%%%%%%%%%%%%%%%%%% INTRODUCTION %%%%%%%%%%%%%%%%%%%%%

\section{Introduction}

It is well known, that QCD vacuum generates various quark and gluon condensates,
the vacuum expectation values of quark and gluon fields of nonperturbative origin.
Among them the gluon condensate $\left< {\alpha_s\over\pi} G_{\mu\nu}^a G_{\mu\nu}^a\right>$,
where $G_{\mu\nu}^a$ is the gluon field strength tensor and $\alpha_s={g_s^2\over 4 \pi}$ is the
running QCD coupling constant, plays a special role. The existence of the gluon condensate
in QCD was first demonstrated by Shifman, Vainstein and Zakharov \cite{SVZ}. Its special role is
caused by few reasons. First, it has the lowest dimension $d=4$ among gluon condensates,
as well as any other condensates conserving chirality. For this reason the gluon condensate
is the most important one in determination of hadronic properties by QCD sum rules, if chirality
conserving amplitudes are considered (e.g. in case of meson mass determination). Second, the
value of the gluon condensate is directly related to the vacuum energy density $\ve$.
As was shown in \cite{SVZ}
\be
\ve\,=\, -\,{\pi\over 8\alpha_s^2}\,\beta(\alpha_s)\,\left< {\alpha_s\over\pi} G_{\mu\nu}^a G_{\mu\nu}^a\right>
\ee
where $\beta(\alpha_s)$ is Gell-Mann-Low $\beta$-function. Therefore, the sign and magnitude of
$\left< {\alpha_s\over\pi} G^2\right>$ are very important for theoretical description of QCD vacuum and for
construction of hadron models (e.g. the bag model). Third, in some models the numerical value of
the gluon condensate is usually used as normalization scale, which fixes the model parameters.
For example, in the instanton model it is required, that this value is reproduced
by the model.

The numerical value of the gluon condensate
\be
\label{svzval}
\left< {\alpha_s\over\pi} G_{\mu\nu}^a G_{\mu\nu}^a\right>\,=\,0.012 \, {\rm GeV}^4
\ee
has been found in \cite{SVZ} from charmonium sum rules. (This value is often referred to as the standard
or SVZ value.) Later there were many attempts to determine the gluon condensate by considering various processes
within various approaches. In some of them the value (\ref{svzval}) (or ones, by a factor of $1.5$ higher)
was confirmed \cite{RRY}--\cite{MO}, in others it was claimed,
that the actual value of the gluon condensate is by a factor
2--5 higher than (\ref{svzval}) \cite{B}--\cite{NR0}.

From today's point of view the calculations performed in \cite{SVZ} have a serious drawback.
Only the first order (NLO) perturbative correction was accounted in \cite{SVZ} and it was taken
rather low value of $\alpha_s$, later not confirmed by the experimental data.
(It was assumed, that QCD parameter $\Lambda^{(3)}\approx 100\, {\rm MeV}$ and $\alpha_s(m_c)\approx 0.2$,
today's values are essentially higher.) The contribution of the next, dimension 6, operator
$G^3$ was neglected, so the convergence of the operator product expansion was not tested.
In charmonium sum rules the moments $M_n(Q^2)$ of the polarization function $\Pi(q^2)$, $q^2=-Q^2$
were calculated at the point $Q^2=0$. It was shown in \cite{NR0}, that the higher order terms of
the operator product expansion (OPE), namely the contributions of $G^3$ and $G^4$ operators are of
importance at $Q^2=0$. The results of calculations
 of the second order (NNLO) perturbative corrections to $\Pi(q^2)$ as well as $\alpha_s$-correction
 to the gluon condensate are available now. They demonstrate, that both of them as a rule are large
 and by no means can be neglected in the sum rules for the moments at $Q^2=0$.
  Finally, the experimental data shifted significantly in comparison with ones, used in \cite{SVZ}.

Later the charmonium sum rules were considered at the NLO level in \cite{RRY} for $Q^2 > 0$
and their analysis basically confirmed the results of \cite{SVZ}.  There are recent publications
\cite{JP},  \cite{EJ}, \cite{KS} where the charmonium as well as bottomonium sum rules were
analyzed at $Q^2=0$ with $\alpha_s^2$ perturbative corrections in order to extract
the charm and bottom quark masses in various schemes. The condensate is usually taken
 to be 0 or some another fixed value. However, the charm mass and the condensate
values are entangled in the sum rules. This can be easily understood for large $Q^2$, where
the mass and condensate corrections to the polarization operator behave as some series
in negative powers of $Q^2$, and one may eliminate the condensate contribution to a great
extent by slightly changing the quark mass.  Vice versa, different condensate values
may vary the charm quark mass within few percents.

The condensate could be also determined from other sum rules, which do not involve the
charm quark mass, but the accuracy usually appears to be rather low for this purpose.
In particular, precise analysis of $e^+e^-$ data  \cite{EJKV} lead only to rather weak restrictions
on the gluon condensate. In ref \cite{GIZ} the thorough analysis of
hadronic $\tau$-decay structure functions was performed and the restriction
$\left<{\alpha_s\over\pi}G^2\right>=0.006\pm 0.012 \,{\rm GeV}^4$ was found. This value,
however, does not exclude zero value of the condensate.

 For all these reasons a reconsideration of the problem is necessary.  The
 charmonium sum rules on the next level of precision in comparison with \cite{SVZ} is presented
 below. In Section 2 general outline of the method is given and the experimental input data for the sum rules
 are presented. In Section 3 the method of calculation of the perturbative part of the moments is
 exposed with the references to the sources, we used in the calculations. Section 4 presents the
 gluon condensate contribution with $\alpha_s$-corrections in the form, convenient for
 numerical evaluation of the moments for nonzero $Q^2$. In section 5 perturbative and operator
 product expansion of the moments is considered.
It is argued, that the choice of pole charm quark mass
 as  a mass parameter is not suitable, since in this case the higher order in $\alpha_s$ terms overwhelm
 the lower ones and the $\alpha_s$-series are divergent. It is proposed to get rid of this problem
 by using the $\ov{\rm MS}$ mass as the mass parameter. In what follows the $\ov{\rm MS}$ charm
 quark mass ${\bar m}({\bar m})$ at the renormalization point equal to the mass itself  is
 used. The formulae for moments ${\bar M}_n(Q^2)$, expressed through the $\ov{\rm MS}$ mass, are given
 and the domain in $(n,Q^2)$ plane was found by direct  calculation, where the perturbative series
 are well convergent. In Section 6 the calculation of ${\bar m}({\bar m})$ and gluon condensate is presented.
In Section 7 the sensitivity of the results to the $G^3$ operator contribution is tested.
Section 8 is devoted to the discussion of the attempts to sum up the Coulomb-like corrections.
Section 9 contains the conclusion.

%%%%%%%%%%%%%%% EXPERIMENTAL CURRENT CORRELATOR %%%%%%%%%%%%%%%%%%%%%

\section{Experimental current correlator}

Consider the 2-point correlator of the vector charm currents:
\be
i\int dx \, e^{iqx} \left< \,TJ_\mu (x) J_\nu (0) \, \right>\,=\, (\,q_\mu q_\nu - g_{\mu\nu}q^2
\,)\, \Pi(q^2) \; , \qquad J_\mu = \bar{c} \gamma_\mu c
\label{podef}
\ee
The polarization function $\Pi(q^2)$ can be reconstructed by
 its imaginary part with the help of the dispersion relation:
\be
R_c(s)\,=\,4\pi \, {\rm Im} \, \Pi(s+i0) \; , \qquad
\Pi(q^2)\,=\,{q^2\over 4\pi^2}\int_{4m^2}^\infty \,{R_c(s)\,ds\over s(s-q^2)} \; .
\ee
We shall use notation $R_c$ not to confuse with frequently used notation $R$ for the imaginary
part of the electromagnetic current correlator, $R(s)=\sum_f 3Q_f^2 R_f(s)$,
the normalization $R_c(\infty)=1$ in the parton model.
In the narrow-width approximation $R_c(s)$ can be represented as the sum of the
resonance $\delta$-functions:
\be
\label{rexp}
R_c(s)\,=\,{3 \, \pi \over Q_c^2 \, \alpha_{\rm em}^2\!(s)}\,
\sum_\psi m_\psi \Gamma_{\psi \to ee}\,\delta(s-m_\psi^2)
\ee
where $Q_c=2/3$ is electric charge of $c$-quark, $\alpha_{\rm em}(s)$
is the running electromagnetic coupling:
\be
\alpha_{\rm em}(s)={\alpha(0)\over 1-\Delta\alpha(s)} \; , \qquad
\Delta\alpha(s)=-4\pi\alpha(0)\Pi_{\rm em}(-s)=
\Delta\alpha_{\rm lep}(s)+\Delta\alpha_{\rm had}(s) \; .
\ee
Here $\alpha(0)=1/137.04$ is the fine structure constant, $\Pi_{\rm em}(s)$ is the correlator of
electromagnetic currents $J^{\rm em}_\mu=\sum_i Q_i\bar{\psi}_i\gamma_\mu\psi_i$ defined in
the same way as (\ref{podef}). As usual, the leptonic contribution to $\Pi_{\rm em}(s)$ is found by the
perturbation theory, while the hadronic contribution has to be determined by numerical integration
of experimental $e^+e^-$ (or $\tau$-decay) data. Since $\alpha_{\rm em}(s)$ weakly changes from one resonance
to another, we fix it at $s=m_{J/\psi}^2$ from now on:
$$
\Delta\alpha_{\rm lep}(m_{J/\psi}^2)=0.016 \; , \qquad
\Delta\alpha_{\rm had}(m_{J/\psi}^2)=0.009 \; , \qquad
\alpha_{\rm em}(m_{J/\psi}^2)= 1/133.6
$$

There are 6 vector charmonium states with $J^{PC}=1^{--}$ \cite{PDG}:
$$
\ba{llll}
{\rm notation} \quad & \mbox{mass, MeV} & \mbox{full width, MeV} & \Gamma_{\psi\to ee}, \; {\rm keV} \\
\hline
J/\psi(1S)  &  3096.87\pm 0.04 \quad & 0.087\pm 0.005  &  5.26 \pm 0.37 \\
\psi(2S)     &  3685.96\pm 0.09 & 0.300\pm 0.025  &  2.19\pm 0.15 \\
\psi(3770) &  3769.9\pm 2.5 & 23.6\pm 2.7 & 0.26\pm 0.04 \\
\psi(4040) &  4040\pm 10 &  52\pm 10 & 0.75\pm 0.15 \\
\psi(4160) &  4159\pm 20 & 78\pm 20 & 0.77\pm 0.23 \\
\psi(4415) &  4415\pm 6 &  43\pm 15 & 0.47\pm 0.10 \\
\hline
\ea
$$
The first two resonances, $J/\psi$ and $\psi(2S)$, are sufficiently narrow and their
contribution to $R_c(s)$ can be well parametrized by the $\delta$-functions (\ref{rexp}).

\begin{figure}[tb]
\hspace{35mm}\epsfig{file=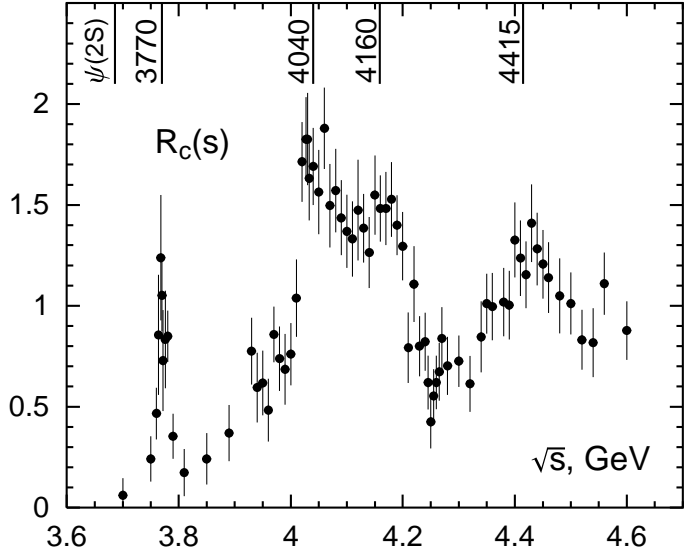, width=90mm}
\caption{$R_c(s)$ in the region of high resonances, determined from BES data \cite{BES}}
\label{rc}
\end{figure}

But the next resonances, especially the last three ones, are rather wide and the narrow-width approximation
for them could be inaccurate. Here it is better to use $R_c(s)$, extracted from $e^+e^-\to hadrons$
branching ratio $R(s)=\sum_f 3Q_f^2 R_f(s)$, which is measured experimentally in wide range of $s$.
Precise data on $R(s)$ in the region  of high charmonium states were obtained recently by BES collaboration
\cite{BES}. In order to extract $R_c(s)$ from these data, one has to subtract the contribution of the
light quarks from $R(s)$. We suppose, that it is well described by the perturbative QCD, which gives $2.16$.
The result for $R_c(s)$ is shown in Fig \ref{rc}. Above the last resonance $R_c(s)$ is getting close to 1,
the parton model prediction.

Now we summarize the following experimental input for $R_c(s)$, which will be used in our
calculations:
\be
\ba{ll}
 \hspace{9mm} s< s_1 = (3.7\, {\rm GeV})^2: &
 \mbox{$\delta$-functions from $J/\psi$ and $\psi(2S)$ according to eq (\ref{rexp})} \\
 s_1<s<s_2=(4.6 \, {\rm GeV})^2: & \mbox{BES data, see Fig \ref{rc}} \\
 s_2<s: & \mbox{continuum, $R_c(s)=1$}
\ea
\label{exinp}
\ee
One could include the $\alpha_s$-correction in the continuum region, but they will not be essential in
what follows.

In order to suppress the contribution of the high energy states, one considers the derivatives
of the polarization function in euclidean region $q^2=-Q^2<0$, the so-called moments:
\be
\label{momdef}
M_n(Q^2)  \equiv {4\pi^2\over n!} \left( - {d\over dQ^2}\right)^n\Pi(-Q^2)=
\int_0^\infty {R_c(s)\, ds\over (s+Q^2)^{n+1}}
\ee
The experimental values are calculated according to (\ref{exinp}):
\be
\label{momexp}
M_n(Q^2)\,=\,{27\,\pi\over 4\, \alpha_{\rm em}^2}\sum_{\psi=1}^2
{m_\psi\Gamma_{\psi\to ee}\over (m_\psi^2+Q^2)^{n+1}}
\,+\,\int_{s_1}^{s_2} {R_c(s)\, ds\over (s+Q^2)^{n+1}}
\,+\,{1\over n (s_2+Q^2)^n}
\ee
The squared error of the moments (\ref{momexp})
is computed as the sum of the squared errors of each term.

The lowest state $J/\psi$ gives maximal contribution to the moments due to the largest
width $\Gamma_{J/\psi\to ee}$, which itself has the error $7\%$. This error can be
eliminated to a great extent, if one considers the ratio of two moments, which in general
case can be written in the following form:
\be
\label{momra}
r(n_1,n_2; Q^2) \, \equiv \, {M_{n_1}(Q^2)\over M_{n_2}(Q^2)}
\,=\,(m_{J/\psi}^2+Q^2)^{n_2-n_1} \, {1+D_1\over 1+D_2} \; ,
\ee
where $D_{1,2}$ denote the higher state contribution to the moments (\ref{momexp})
divided by the $J/\psi$ contribution. Then the error of this ratio is calculated by usual rules:
\be
\label{momraer}
\left({\Delta r\over r}\right)^2=\sum_{j=1}^2 \left( {\Delta D_j\over 1+D_j} \right)^2 \; ,
 \ee
where the mass errors are neglected. If $D_{1,2}\ll 1$, the relative error of the ratio
is much smaller than the relative errors of the moments itself. This fact has been utilized in
many papers on charmonium sum rules and will be used here.

In our calculations we shall always use sufficiently high $n$ ($n\ge 8$), so that
the last term in (\ref{momexp}), which comes from continuum,
is small compared to the resonance contribution and the uncertainty
introduced by this term is negligible. Moreover, the difference between the narrow width approximation
for the high resonances (above $\psi(2S)$) as given by (\ref{rexp}), and their representation
by (\ref{exinp}) is small and well below the quoted errors.

%%%%%%%%%%%%%%%% THEORETICAL R(s) %%%%%%%%%%%%%%%%%%%%%%%%%%

\section{Theoretical {\boldmath $R(s)$}}

At first one defines the running QCD coupling $a(\mu^2)\equiv\alpha_s(\mu^2)/\pi$
 as a solution of the renormalization group equation:
\be
\label{rgas}
\int_{a(\mu_0^2)}^{a(\mu^2)}{da\over \beta(a)}\,=\,-\ln{\mu^2\over \mu_0^2} \; , \qquad
\beta(a)\,=\,\sum_{n\ge 0}\beta_n a^{n+2}
\ee
Then the functions $R^{(n)}(s,\mu^2)$ are defined as the coefficients in the $\alpha_s$-expansion:
\be
\label{rcf}
R_c(s)\,=\,\sum_{n\ge 0} R^{(n)}(s,\mu^2)\, a^n(\mu^2)
\ee
Since $R_c(s)$ is the physical quantity, it does not depend on the scale $\mu^2$, although each term in
(\ref{rcf}) may be $\mu^2$ dependent.

It is easier to represent the results in terms of the pole quark mass $m$ and the velocity
$v=\sqrt{1-4m^2/s}$. The first two terms in the expansion (\ref{rcf}) do not depend on $\mu^2$.
The leading term $R^{(0)}$ was calculated in \cite{BP}, the next to leading
$R^{(1)}$ in \cite{Schw}:
\bea
R^{(0)} & = &  {v\over 2}(3-v^2)  \nonumber \\
R^{(1)} & = & {v\over 2}(5-3v^2)+2v\,(3-v^2)\left( \ln{1-v^2\over 4} -{4\over 3} \ln{v} \right) + {v^4\over 3}
\ln{1+v\over 1-v} +{4\over 3} (3-v^2)(1+v^2) \nonumber \\
 & & \times
\left[ \,2\,{\rm Li}_2\!\left( {1-v\over 1+v} \right) +\,{\rm Li}_2\!\left( -{1-v\over 1+v} \right)
+ \left( {3\over 2}\ln{1+v\over 2}- \ln{v}+{11\over 16} \right) \ln{1+v\over 1-v}\,\right]
\label{rcf12}
\eea
where ${\rm Li}_2(x)=\sum_{n=1}^\infty {x^n\over n^2}$ is the dilogarithm function.
The function $R^{(2)}$ is usually decomposed into four gauge invariant terms:
\be
\label{r2parts}
R^{(2)}\,=\,C_F^2 R^{(2)}_A \,+\, C_A C_F R^{(2)}_{NA} \,+\, C_F T n_l R^{(2)}_l\, +\, C_F T R^{(2)}_F \; ,
\ee
where $C_A=3$, $C_F=4/3$,  $T=1/2$ are group factors, $n_l=n_f-1$ is the number
of light quarks. The function $R^{(2)}_l$ comes from the diagram with two quark loops:
one loop with massive quark, which couples to the vector currents,
 and another massless quark loop (the so-called double bubble diagram).
It was originally found in \cite{HKT} and in our normalization takes the form:
\be
R^{(2)}_l\,=\,\left( \,-\,{1\over 4}\,\ln{\mu^2\over 4s}\, -\, {5\over 12}\, \right) R^{(1)}\,+\,\delta^{(2)}
\ee
where the function $\delta^{(2)}$ is given by equation (B.3) in ref \cite{CKS}. The function
$R^{(2)}_F$ comes from the similar double bubble diagram with equal quark masses and has the
form \cite{CHKST}:
\be
R^{(2)}_F\,=\,\rho^V\,+\,\rho^R\,-\,{1\over 4}\,R^{(1)}\,\ln{\mu^2\over m^2}
\ee
where $\rho^V$ is given by equation (12) in ref \cite{CHKST}. The function $\rho^R$
comes from the 4-particle cut and vanishes for $s<16\, m^2$. It is represented as the
double integral (13) in ref \cite{CHKST} which can be computed numerically.
However for $s>16\, m^2$ the total function $R^{(2)}_F$ can be well approximated
by its high energy asymptotic:
\be
R^{(2)}_F\,=\,-\,{1\over 4}\,R^{(1)}\,\ln{\mu^2\over s}\,+
\,\zeta_3\,-\,{11\over 8}\,-\,{13\over 2}\,{m^2\over s} \, +\, O\!\left(m^4/s^2\right)
\ee
In numerical calculations we take all the terms up to $m^{12}/s^6$, extracted from \cite{CHKS}.
The functions $R^{(2)}_A$ and $R^{(2)}_{NA}$ are generated by the diagrams with single quark loop
and various gluon exchanges, $R^{(2)}_{A}$ is abelian part while $R^{(2)}_{NA}$ contains purely
nonabelian contributions. They are not known analytically. We will
use the approximations, given by equations (65), (66) in ref \cite{CKS}
(divided by 3 in our conventions) which reproduce all known asymptotics
and Pade approximations with high accuracy.

%%%%%%%%%%%%%%%%%%%% CONDENSATE CONTRIBUTION %%%%%%%%%%%%%%%%%%%%

\section{Condensate contribution}

 The contribution of the dimension 4
gluon condensate $\left< aG^2\right>\equiv \left< {\alpha_s\over \pi}G_{\mu\nu}^a G_{\mu\nu}^a\right>$
to the polarization function of massive quarks has the form:
$$
\Pi^{(G)}(-Q^2)\, = \, {\left< aG^2\right> \over (4m^2)^2}
\left[\,f^{(0)}(z)\,+\,a\,f^{(1)}(z)\,\right]  \; ,\qquad
\mbox{where} \qquad z={-Q^2\over 4m^2} \; ,
$$
The leading order function was found in \cite{SVZ}:
\be
f^{(0)}(z)\,=\,-\,{1\over 12\, z^4 v^4} \,\left[ \,{3\over 8} \,{2z-1\over z v} \, \ln{v-1\over v+1} \,+ \,
 z^2\,-\,z\,+\,{3\over 4} \,\right] \; ,
\ee
where $v=\sqrt{1-1/z}$.
For this function the following dispersion-like relation can be written:
\be
\label{drcf0}
f^{(0)}(z)\,=\, -\,{1\over 12} \int_1^\infty {dz'\over {z'}^3 v'}
 \left[ \,{3\over 4}\,{1\over (z'-z)^2}\,+\,{z'\over (z'-z)^3}\,\right]
\ee
This representation is convenient for an evaluation of various transformations of the
polarization function $\Pi(s)$, in particular, the moments.

The next-to-leading order function $f^{(1)}$ was explicitly
found in \cite{BBIFTS}.  One could differentiate it $n$ times to obtain the moments
for arbitrary $Q^2$. However, we prefer to construct the
dispersion integral similar to (\ref{drcf0}).
The function $f^{(1)}(z)$ has a cut from $z=1$ to $\infty$ and behaves as $v^{-6}$ at $z\to 1$.
Integrating $f^{(1)}(z')/(z'-z)$ by $z'$ along the contour around the cut, one obtains the following
representation:
 \bea
 f^{(1)}(z) & = & {1\over \pi}\int_{1+\eps}^\infty  {{\rm Im}\, f^{(1)}(z'+i0)\over z'-z} \,dz' \,
+\,\sum_{i=1}^3 {\pi^2f_i\over (1-z)^i} \,-\, {65\over 1152}\,{\eps^{-3/2}\over 1-z}
\nonumber \\
 & & + \left[ \,{8633 \over 6912}\,+\,{17 \over 36}\ln{(8\eps)}\,\right] {\eps^{-1/2}\over 1-z}
\,+\, {65 \over 384}\, {\eps^{-1/2} \over (1-z)^2}
\label{cif1}
\eea
where $\eps\to 0$ and
\be
\label{cfdr1fi}
f_1\,=\,-\,{17\over 384} \; , \qquad f_2\,=\, -\, {413\over 6912}\; , \qquad f_3\,=\, -\, {197\over 2304}
\ee
The imaginary part is:
\be
\label{drcf1}
 {\rm Im}\, f^{(1)}(z+i0)= {\pi\over 96 z^5 v^5}\left[ P_2^V(z) + {P_3^V(z)\over zv} \ln{1-v\over 1+v}
 +P_4^V(z)(1-z)\left(2\ln{v}+{3\over 2}\ln{(4z)}\right)\right]
\ee
where the polynomials $P_i^V(z)$ are given in the Table 1 of ref \cite{BBIFTS}.
It behaves as $v^{-5}$ at $z\to 1$, so the integral in (\ref{cif1}) is divergent in the limit
$\eps\to 0$. We decompose it into 3 parts
\be
{1\over \pi} {\rm Im}\, f^{(1)}(z+i0) \,=\, F_1(z)\,+\,F_2'(z)\,+\,{1\over 2}\,F_3''(z)
\ee
in such way, that each function $F_i(z)$ behaves as $v^{-1}$ at $z\to 1$
and has appropriate asymptotic at $z\to\infty$. In particular we choose
\bea
 F_1(z) & = &  {1\over 96 \,z^4 v}\left[
 {1627\over 36} + {893\over 9}z - {98 \over 3} z^2 + 368 z^3 \right. \nonumber \\
 & & + \left({89\over 6} -22z - {140 \over 3}z^2 + {3208\over 3}z^3 - 3232 z^4 + 2208 z^5\right)
{1\over zv} \ln{1-v\over 1+v} \nonumber \\
 & & + \left. \left( 68z+ {1688 \over 9} z^2 - {4256 \over 3}z^3 + 1472 z^4\right)
 \left(2\ln{v}+{3\over 2}\ln{(4z)}\right)\right] \nonumber \\
F_2(z) & = &  {1\over 96 \,z^4 v}\left[
 -{23\over 4} + {559\over 18}z + {272 \over 3} z^2  \right.
  + \left(-{197\over 8} + {947 \over 72}z\right) {1\over zv} \ln{1-v\over 1+v} \nonumber \\
 & & + \left. {136 \over 3} z^2\left(2\ln{v}+{3\over 2}\ln{(4z)}\right)\right] \nonumber \\
F_3(z) & = &  {1\over 96 \,z^3 v}\left[
 -{67\over 6} - {197\over 12\, zv} \ln{1-v\over 1+v} \right]
\label{imfi}
 \eea
Then one may integrate (\ref{cif1}) by parts twice
and all singular in $\eps$ term cancel. Eventually we obtain
following representation for the function $f^{(1)}$:
\be
f^{(1)}(z)  = \sum_{i=1}^3 \left[ \,{\pi^2f_i\over (1-z)^i}\,
  + \, \int_1^\infty {F_i(z')\over (z'-z)^i}\,dz'\, \right]
\label{cf1dr}
\ee
It will be used to compute the moments numerically.

%%%%%%%%%%%%%%%%%%%% MOMENTS %%%%%%%%%%%%%%%%%%%%%%%

\section{Moments in  {\boldmath $\ov{\rm MS}$} scheme}

For definiteness let us choose the scale $\mu^2=m^2$ in (\ref{rcf}) and write down the
$\alpha_s$-expansion of the moments (\ref{momdef}):
\be
\label{mcf}
M_n(Q^2)\,=\,\sum_{k\ge 0} M_n^{(k)}(Q^2)\, a^k(m^2) \, + \, \left<{\alpha_s\over\pi}G^2\right>
\,\sum_{k\ge 0} M_n^{(G,k)}(Q^2)\, a^k(m^2)
\ee
The perturbative coefficient functions are:
\be
\label{mcfe}
M_n^{(k)}(Q^2)\,=\,\int_{4m^2}^\infty {R_n^{(k)}(s,m^2) \, ds\over (s+Q^2)^{n+1}}
\ee
The leading order can be expressed via Gauss hypergeometric function:
\be
\label{mnlo}
 M_n^{(0)}(Q^2) \, = \, {1\over (4m^2)^n} \, {3\sqrt{\pi}\over 4}\, {(n+1)\Gamma(n)\over \Gamma(n+5/2)}
 \,{}_2\!F_1\!\left( \left.{n,\,n+2\atop n+5/2}\right| -\,{Q^2\over 4m^2}  \right)
\ee
The higher order functions $M^{(1)}$ and $M^{(2)}$ are computed numerically by (\ref{mcfe}).
(Notice, that the analytical expression for $M_n^{(1)}(0)$ has been found in \cite{JP}
and the first 7 moments $M_n^{(2)}(0)$ can be determined from the low energy
expansion of the polarization function $\Pi(s)$ available in \cite{CKS}.)

The leading order contribution of the gluon condensate
is easily obtained from (\ref{drcf0}):
 \be
 M_n^{(G,0)}(Q^2)\,=\, -\,{\pi^2 \over (4m^2)^{n+2}}
 \, {\sqrt{\pi}\over 6 }\,{(n+1)\Gamma(n+4)\over \Gamma(n+7/2)}
 \,{}_2\!F_1\!\left( \left.{n+2,\,n+4\atop n+7/2}\right| -\,{Q^2\over 4m^2}  \right)
\ee
The next-to-leading condensate correction
 can be computed numerically with the help of the integral
 representation, obtained from (\ref{cf1dr}):
\be
M^{(G,1)}_n(Q^2)\,=\, {4\pi^2\over (4m^2)^{n+2}}\sum_{i=1}^3
{\Gamma(n+i)\over\Gamma(n+1)\Gamma(i)}\left[\, {\pi^2 f_i \over (1+y)^{n+i}}\,
 + \, \int_1^\infty {F_i(z)\over (z+y)^{n+i}}\,dz\, \right] \; ,
 \label{mc1c}
\ee
where $y=Q^2/(4m^2)$, the constants $f_i$ and the functions $F_i(z)$ are given in
(\ref{cfdr1fi}) and (\ref{imfi}).

The pole quark mass $m$ is the most natural choice, since it is the physical invariant.
However in the pole scheme the perturbative corrections to the moments are huge.
For instance, at the typical point, which will be used later in our analysis, one gets:
\be
n=10 \; , \; Q^2=4m^2 :  \qquad
{M^{(1)}\over M^{(0)}}=13.836 \; , \qquad
{M^{(2)}\over M^{(0)}}=193.33 \; , \qquad
{M^{(G,1)}\over M^{(G,0)}}=13.791
\label{n10q1p}
\ee
Since in the domain of interest $a\sim 0.1$, this is an indication,
that the series (\ref{mcf}) is divergent. The situation is even worse
for $Q^2=0$ (see \cite{JP}). It is almost impossible to choose an informative region in the
$(n,Q^2)$ plane where the perturbative corrections in the pole mass scheme are tolerable and
the continuum as well as $\left<G^3\right>$ contributions are suppressed enough on the other hand.

The traditional solution to this problem is the mass redefinition. In particular, in the most
popular $\ov{\rm MS}$ scheme the mass corrections are known to be significantly smaller.
In $\ov{\rm MS}$ conventions the mass $\bar{m}$ depends on the scale $\mu^2$ according to
the RG equation:
\be
 {\bar m}(\mu^2)={\bar m}(\mu_0^2)\exp \left(
 \int_{a(\mu_0^2)}^{a(\mu^2)} {\gamma_m(a)\over \beta(a)}da \right)\; , \qquad
 \gamma_m(a) = \sum_{n\ge 0} \gamma_n a^{n+1} \; ,
 \label{msmrun}
 \ee
 where $\gamma_m$ is the mass anomalous dimension. In what follows we shall
choose the most natural mass scale $\mu^2={\bar m}^2$ and will denote ${\bar m}({\bar m}^2)$
as simply ${\bar m}$.

There is a perturbative relation between the pole mass $m$ and $\ov{\rm MS}$ one ${\bar m}$:
\be
{m^2\over {\bar m}^2} \,=\, 1\,+\,\sum_{n\ge 1} \, K_n \,a^n({\bar m}^2)
\label{msbvsp}
\ee
The 2-loop factor was found, in particular, in \cite{GBGS}, while the 3-loop
factor was recently calculated in \cite{MvR}:
\bea
 K_1& =&  {8\over 3}  \nonumber \\
 K_2& =&  28.6646 - 2.0828\,n_l \, = \,  22.4162 \nonumber \\
 K_3&=&  417.039 - 56.0871\, n_l + 1.3054 \, n_l^2 \, = \, 260.526
 \label{kfacnum}
 \eea
 We put $n_l=3$ in the last column. The series (\ref{msbvsp}) also looks divergent at the
charm scale. (Notice, that the authors of \cite{SVZ} used another mass convention, although numerically
close to $\ov{\rm MS}$ scheme at the NLO level: the coefficient $K_1$ was equal to $4\ln{2}$ there.)

\begin{figure}[tb]
\epsfig{file=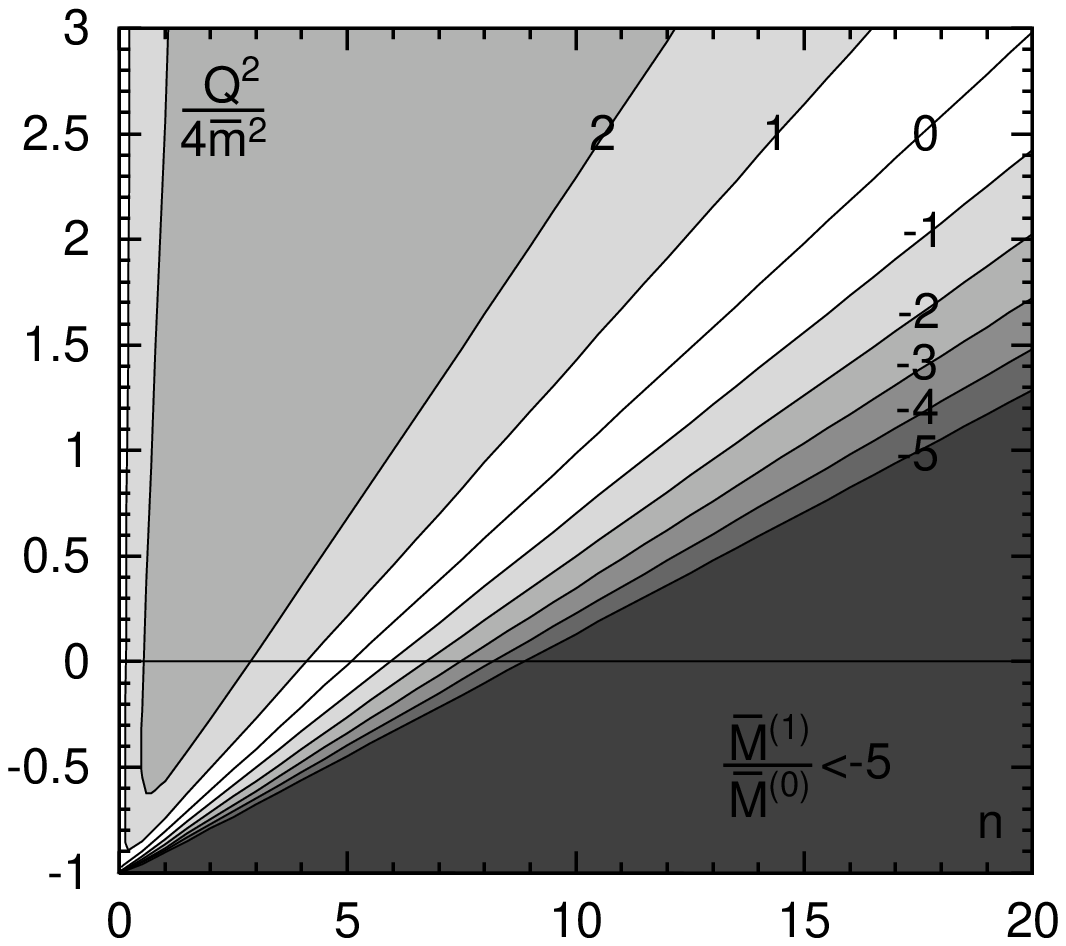, width=82mm}\epsfig{file=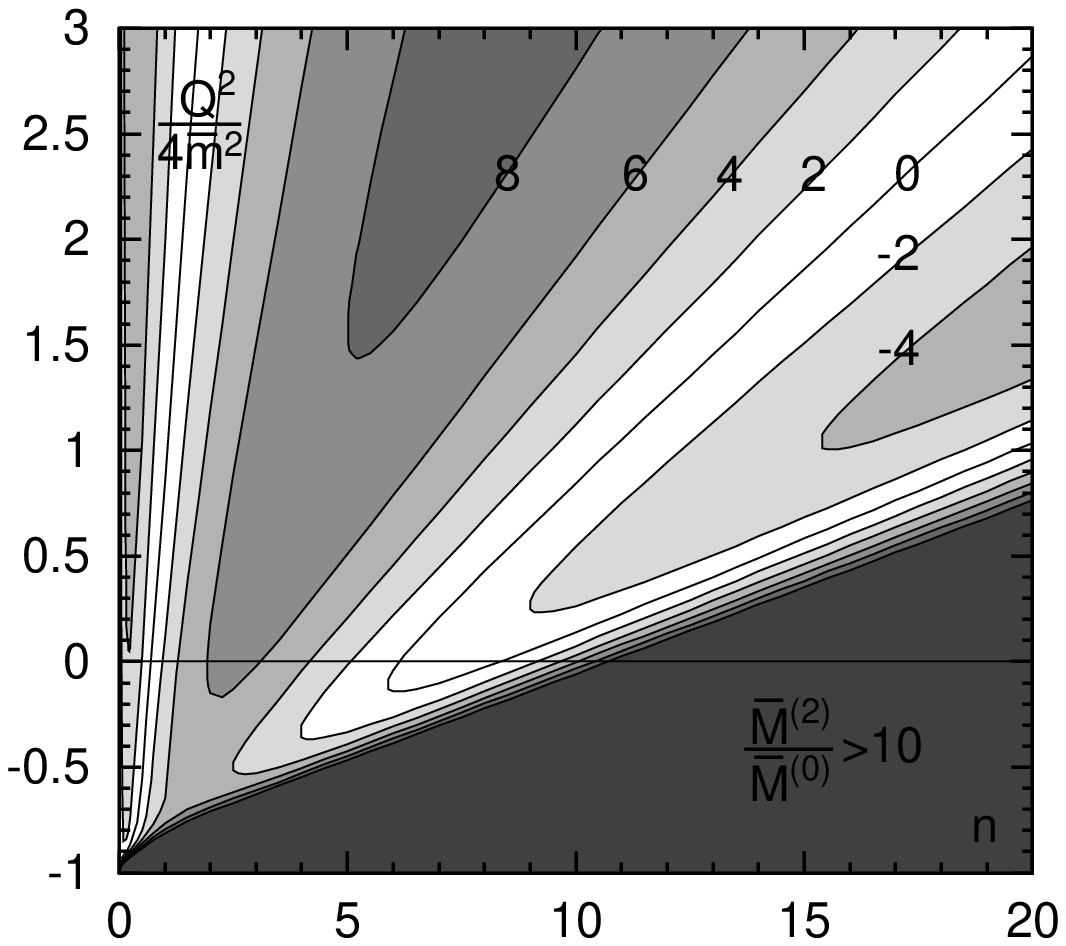, width=82mm}
\caption{Ratio ${\bar M}^{(1)}_n/{\bar M}^{(0)}_n$ (left) and  ${\bar M}^{(2)}_n/{\bar M}^{(0)}_n$ (right)
 in the plane ($n$, $Q^2$). }
\label{fig_mdm}
\end{figure}

Nevertheless let us assume for a moment, that $\alpha_s$ is small, take advantage of (\ref{msbvsp},\ref{kfacnum})
and  express the moments  (\ref{momdef}) in terms of the mass $\bar{m}$:
 \be
\label{maemsb}
M_n(Q^2)\,=\,\sum_{k\ge 0} {\bar M}_n^{(k)}(Q^2)\, a^k({\bar m}^2) \, + \,
\left<{\alpha_s\over\pi}G^2\right> \sum_{k\ge 0} {\bar M}_n^{(G,k)}(Q^2)\, a^k({\bar m}^2)
\ee
As follows from the definition (\ref{momdef}) and dimensional consideration
\bea
{\bar M}_n^{(0)}(Q^2) & = & M_n^{(0)} \nonumber \\
{\bar M}_n^{(1)}(Q^2) & = & M_n^{(1)}-K_1 (n-d/2)\, M_n^{(0)}+K_1(n+1)\,Q^2M_{n+1}^{(0)} \nonumber \\
{\bar M}_n^{(2)}(Q^2) & = & M_n^{(2)}-K_1 (n-d/2)\, M_n^{(1)}+K_1(n+1)\,Q^2M_{n+1}^{(1)}\nonumber \\
 & & + \, (n-d/2) \left[ {K_1^2\over 2} (n+1-d/2) - K_2 \right]  M_n^{(0)}  \nonumber \\
 & & + \, (n+1) \left[ K_2  -K_1^2(n+1-d/2)\right] Q^2M_{n+1}^{(0)}
  + \, {K_1^2\over 2} (n+1)(n+2) \,Q^4 M_{n+2}^{(0)} \nonumber\\
{\bar M}_n^{(G,0)}(Q^2) & = & M_n^{(G,0)} \nonumber \\
{\bar M}_n^{(G,1)}(Q^2) & = & M_n^{(G,1)}-K_1 (n+2-d/2) \, M_n^{(G,0)} +
  K_1(n+1)\,Q^2M_{n+1}^{(G,0)}
\label{mommsb}
\eea
where $d$ is the dimension of the polarization function $\Pi(Q^2)$ ($d=0$ for vector currents),
all $M_n^{(i)}$ in the rhs are computed with $\ov{\rm MS}$ mass ${\bar m}$.

\begin{figure}[tb]
\hspace{41mm}\epsfig{file=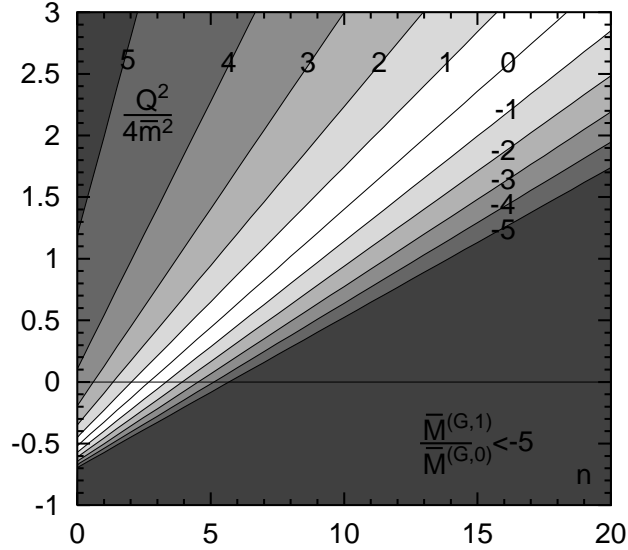, width=82mm}
\caption{Ratio ${\bar M}^{(G,1)}_n/{\bar M}^{(G,0)}_n$ in the plane ($n$, $Q^2$)}
\label{fig_mgdmg}
\end{figure}

The moment corrections ${\bar M}^{(k)}$ are much smaller than $M^{(k)}$ in the pole scheme.
In particular, at the same point, which was considered in (\ref{n10q1p}), we have now:
\be
n=10 \; , \; Q^2=4{\bar m}^2:  \qquad
{{\bar M}^{(1)}\over {\bar M}^{(0)}}=0.045 \; , \qquad
{{\bar M}^{(2)}\over {\bar M}^{(0)}}=1.136 \; , \qquad
{{\bar M}^{(G,1)}\over {\bar M}^{(G,0)}}=-1.673
\label{n10q1msb}
\ee
This smallness of corrections as compared to the pole scheme is observed for almost all $n$ and $Q^2$.
The ratios   ${\bar M}^{(1)}_n/{\bar M}^{(0)}_n$  and  ${\bar M}^{(2)}_n/{\bar M}^{(0)}_n$
are shown in Fig \ref{fig_mdm} and the ratio ${\bar M}^{(G,1)}_n/{\bar M}^{(G,0)}_n$ in
the Fig \ref{fig_mgdmg} for $n=0\ldots 20$ and $Q^2/(4{\bar m}^2)= -1\ldots 3$.
The perturbative expansion in $\ov{\rm MS}$-scheme
obviously does not work in the area of high $n$ and low $Q^2$, marked with dark.
(The detailed data are presented in the Tables 1,2,3 in the Appendix.)

Now we can argue, why the expression (\ref{mommsb}) for the moments is legitimate, despite that the series
(\ref{msbvsp}), relating the pole mass $m$ and $\ov{\rm MS}$ mass ${\bar m}$, is divergent at
the coupling $\alpha_s$ taken on the charm mass scale. If $\alpha_s$ is small enough, eq (\ref{mommsb})
is correct. In this case the same values of ${\bar M}_n$ can be obtained by the procedure, when $\ov{\rm MS}$
mass renormalization is performed directly in the diagrams, without all the concept of the pole mass.
If the pole mass concept is not used, the relations (\ref{msbvsp},\ref{kfacnum}) are irrelevant. These relations
demonstrate only, that the pole mass is an ill defined object in case of charm. The check of selfconsistency
of $M_n^{(k)}$ moments is the convergence of the series (\ref{maemsb}).

If one takes the QCD coupling at some another scale
 $\alpha_s(\mu^2)$, the function $M^{(2)}$ must be replaced by:
\be
\label{m2shift}
a({\bar m}^2) \, \to \,   a(\mu^2)   \; , \qquad
 {\bar M}_n^{(2)}(Q^2) \, \to \, {\bar M}_n^{(2)}(Q^2) \,+
 \,  {\bar M}_n^{(1)}(Q^2)\, \beta_0 \, \ln{ \mu^2\over{\bar m}^2}
\ee
so that the series (\ref{maemsb}) is $\mu^2$-independent at the order $\alpha_s^2$.

%%%%%%%%%%%%%%%%%% DETERMINATION... %%%%%%%%%%%%%%%%%%%%%%%%%%%%

\section{Determination of charm quark mass and gluon condensate from data}

Theoretical moments depend on 3 parameters: charm quark mass, QCD coupling constant and gluon
condensate. The QCD coupling $\alpha_s$ is universal quantity and can be taken from other experiments.
In particular, as boundary condition in the RG equation (\ref{rgas}) we put:
 \be
 \label{alphatau}
 \alpha_s(m_\tau^2) \, =\, 0.330\pm 0.025 \; , \qquad     m_\tau=1.777\,{\rm GeV}
 \ee
found from hadronic $\tau$-decay analysis \cite{GIZ} at the $\tau$-mass
in agreement with other data \cite{PDG}.

Another question is the choice of the scale $\mu^2$, at which $\alpha_s$ should be taken.
Since the higher order perturbative corrections are not known, the moments $M_n(Q^2)$
will depend on this scale. In the massless limit the most natural choice is $\mu^2=Q^2$.
On the other hand for massive quarks and $Q^2=0$ the scale is usually taken $\mu^2\sim m^2$.
So we choose the interpolation formula:
\be
\label{ascale}
\mu^2\,=\,Q^2\,+\,{\bar m}^2
\ee
 At this scale $\alpha_s$ is smaller than at $\mu^2={\bar m}^2$ for the price of larger
 ${\bar M}^{(2)}_n$ according to (\ref{m2shift}). (Notice, that in the Tables in the Appendix as well as
 in the Fig \ref{fig_mdm} the ratio ${\bar M}^{(2)}/{\bar M}^{(0)}$ is given at the scale $\mu^2={\bar m}^2$.)
Sometimes we will vary the coefficient before ${\bar m}^2$  (\ref{ascale}) to test the
dependence of the results on the scale.

The sum rules for low order moments $M_n(Q^2)$, $n\le 3$ cannot be used because of large
contribution of high excited states and continuum as well as large $\alpha_s^2$ corrections (see the
Tables in Appendix), especially at $Q^2=0$. As the Fig \ref{fig_mgdmg} demonstrates,
at $n\ge 4$ the $\alpha_s$ correction to the gluon condensate
is large at $Q^2=0$. The $\left< G^3\right>$ condensate contribution is also large (see below),
which demonstrates, that the operator product expansion is divergent here. For these reasons
we will avoid using the sum rules at small $Q^2$.

\begin{figure}[tb]
\epsfig{file=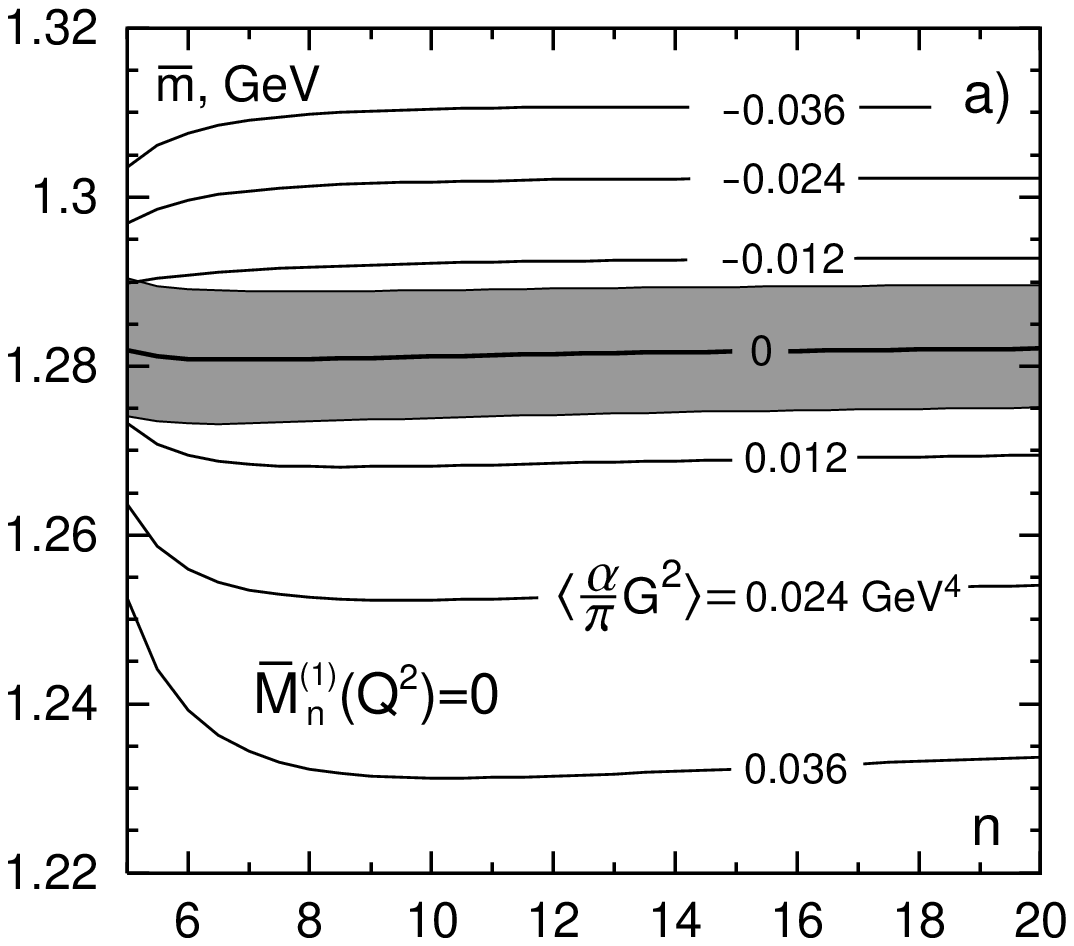, width=82mm} \epsfig{file=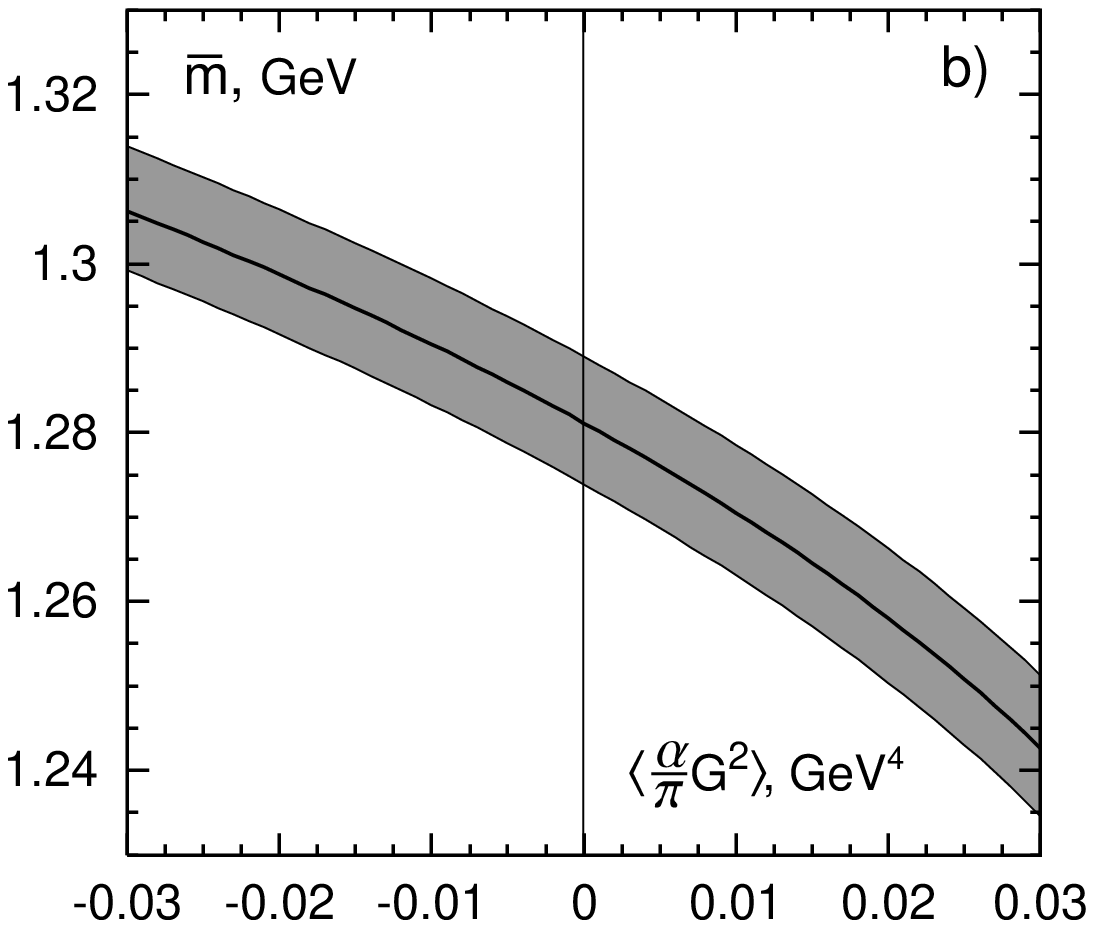, width=82mm}
\caption{a): $\ov{\rm MS}$ mass found from experimental moments $M_n(Q^2_n)$
for different $n$ and $Q^2_n$ determined by the equation ${\bar M}_n^{(1)}(Q^2_n)=0$
for different values of the gluon condensate. The shaded area shows the experimental
error for  $ \left<{\alpha_s\over \pi} G^2\right>=0$, for nonzero condensates only
the central lines are shown. b): ${\bar m}({\bar m}^2)$ in GeV
vs $ \left<{\alpha_s\over \pi} G^2\right>$
 in GeV${}^4$ determined from $n=10$ and $Q^2= 0.98 \times 4{\bar m}^2$.
The $\alpha_s$ is taken at the scale (\ref{ascale}).}
\label{msb}
\end{figure}

As the Fig \ref{fig_mdm} shows, the first correction to the moments ${\bar M}_n^{(1)}(Q^2)$
vanishes along the diagonal line, approximately parametrized by the equation
$Q^2/(4\bar{m}^2)=n/5-1$. The second-order correction ${\bar M}^{(2)}$ and the correction
to the condensate contribution ${\bar M}^{(G,1)}$ are also small along this diagonal for $n>5$.
Now lest us compare the theoretical moments
with experimental value (\ref{momexp}) at different points on this diagonal.
If the condensate is fixed, then one can numerically solve this equation
in order to find the $\ov{ \rm MS}$ mass. The result is shown in Fig \ref{msb}a.
 The values $n<5$ are not reliable, since the $\alpha_s$-correction
 to the condensate exceeds $-50\%$ here.

 The lines in Fig \ref{msb}a are almost horizontal, if the condensate is not too large.
 Consequently there is a correlation between the mass
 and condensate and we establish the dependence of the $\ov{\rm MS}$ charm mass
 ${\bar m}$ on the condensate $ \left<{\alpha_s\over \pi} G^2\right>$
 found  at the point $n=10$, $Q^2=0.98 \times 4{\bar m}^2$ on this diagonal.
 It is plotted in the fig \ref{msb}b. The error of the experimental moments
 is about $7\%$, arising mainly from the uncertainty in $\Gamma_{J/\psi\to ee}$.
 But, since $M_n(Q^2)\sim (4{\bar m}^2+Q^2)^{-n}$, the mass error is of order $7/n\%$, i.e. is much smaller.
For instance, at zero condensate
\be
{\bar m}({\bar m}^2)\,=\,1.283 \pm 0.007 \, {\rm GeV} \qquad \mbox{for} \qquad
\left<{\alpha_s\over \pi} G^2\right>=0
\label{msbm}
\ee
the error is purely experimental. The dependence plotted in fig \ref{msb}b as well as
the value (\ref{msbm}) are weakly sensitive to particular choice of the QCD coupling $\alpha_s$
and the scale $\mu^2$. This is an obvious advantage of nonzero $Q^2$ while
the analysis at $Q^2=0$ leads to significantly higher error \cite{KS}.

\begin{figure}[tb]
\epsfig{file=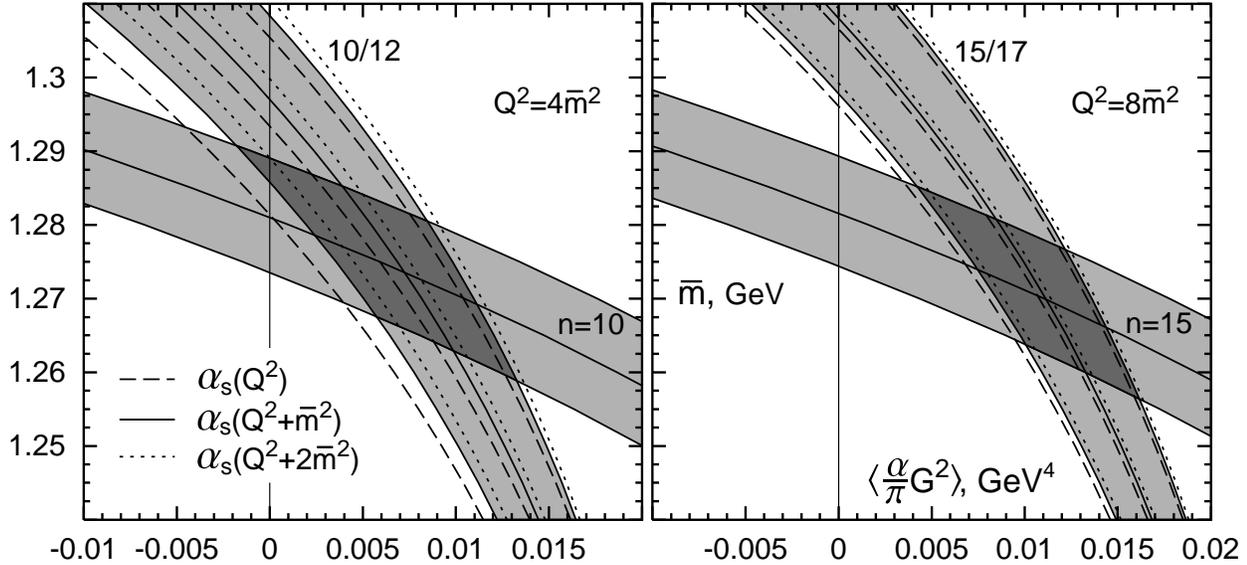, width=164mm}
\caption{$\ov{\rm MS}$ mass versus gluon condensate obtained from
different points on $(n,Q^2)$ plane. "Horizontal" bands obtained from the moments
$(10,4{\bar m}^2)$ and $(15,8{\bar m}^2)$, "vertical" bands obtained from the ratio of the moments
$M_{10}/M_{12}$ (left), $M_{15}/M_{17}$ (right) for  few different choices of $\alpha_s(\mu^2)$.}
\label{mc12}
\end{figure}

It is more difficult to find the restrictions on the mass and condensate separately.
For this purpose one has to choose the point in $(n,Q^2)$ plane which is
1) out of the diagonal, since no new information can be obtained from there,
2) not in the lower right corner (high $n$, low $Q^2$), where perturbative corrections
as well as $\alpha_s$ corrections to the gluon condensate are
large and 3) not in the upper left corner (low $n$, high $Q^2$), where the continuum
contribution to the experimental moments is uncontrollable. It turns out
that if one considers the ratio of the moments (\ref{momra}), the mass--condensate dependence
appears to be different in comparison to the fig \ref{msb}b.
In particular, the results obtained from the ratio $M_{10}/M_{12}$ at $Q^2=4{\bar m}^2$ and
$M_{15}/M_{17}$ at $Q^2=8{\bar m}^2$ are demonstrated in the left and right parts
of the Fig \ref{mc12} respectively. At the same figures the mass-condensate dependence, obtained
from the moments $M_{10}(Q^2=4{\bar m}^2)$ and $M_{15}(Q^2=8{\bar m}^2)$ is also plotted.
The error bands include both the experimental error of the ratio (\ref{momraer})
and the uncertainty of $\alpha_s$ (\ref{alphatau}).
Obviously the results, obtained outside the diagonal, are
sensitive to the choice of $\alpha_s$ as well as $\mu^2$. The small variation of $\mu^2$
slightly changes the acceptable region in the fig \ref{mc12}, but if one takes $\mu^2$ few times lower,
the region expands to the left significantly.

\begin{figure}[tb]
\hspace{41mm} \epsfig{file=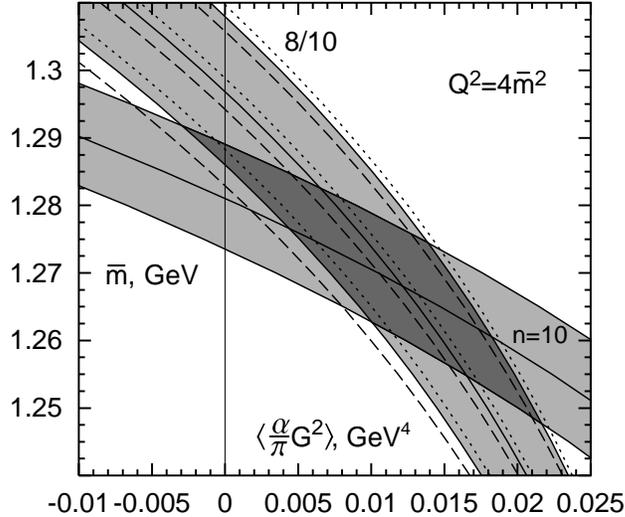, width=82mm}
\caption{$\ov{\rm MS}$ mass versus gluon condensate obtained from
the
ratio $M_8/M_{10}$ above the diagonal. For more notations see Fig \ref{mc12}.}
\label{mc1g3}
\end{figure}

The absolute limits of the $\ov{\rm MS}$ charm quark mass and the gluon condensate
can be determined from the fig \ref{mc12}:
\be
\label{mgrestr}
{\bar m}({\bar m}^2)\,=\,1.275\pm 0.015 \, {\rm GeV} \; , \qquad
\left< {\alpha_s\over \pi} G^2\right>\,=\,0.009\pm 0.007 \, {\rm GeV}^4
\ee
The restrictions on ${\bar m}$ and the gluon condensate, obtained from other ratios of moments,
agree with (\ref{mgrestr}), but are weaker (see Fig \ref{mc1g3}, where the ratio $M_8/M_{10}$ is
considered). The stability intervals in the moments, i.e.~the intervals, where (\ref{mgrestr}) takes place
within the errors, were found to be $n=8-13$ at $Q^2=4{\bar m}^2$ and $n=12-19$ at $Q^2=8{\bar m}^2$.

As a check, the calculations were performed, where the $\alpha_s^2$-terms in the $\ov{\rm MS}$ moments
were omitted as well as the $\alpha_s$-corrections to the gluon condensate contribution
(${\bar M}^{(2)}_n={\bar M}^{(G,1)}_n=0$). At $Q^2=4{\bar m}^2$ it was found ${\bar m}=1.266\,{\rm GeV}$ and
$\left<aG^2\right>=0.011\,{\rm GeV}^4$ from $M_{10}$ and $M_{10}/M_{12}$ while
at $Q^2=8{\bar m}^2$ the result ${\bar m}=1.263\,{\rm GeV}$ and
$\left<aG^2\right>=0.015\,{\rm GeV}^4$ was obtained from $M_{15}$ and $M_{15}/M_{17}$.
These values agree with (\ref{mgrestr}) in the limit of errors. However, it is difficult to estimate the errors
of the calculation, where $\alpha_s^2$ terms are omitted, because of the uncertainty in the scale.

$R_c(s)$ in (\ref{rcf}), or the expression for the moments are, in principle, independent on the normalization
scale $\mu^2$. However, in fact, since we take into account only first 3 terms in the $\alpha_s$-expansion
in (\ref{mcf}), such dependence takes place. Namely, when we change the normalization point in $\alpha_s$
from ${\bar m}^2$ to $\mu^2=Q^2+{\bar m}^2$ (\ref{ascale}) with
the help of eq (\ref{m2shift}), the values of the moments,
defined by (\ref{mcf}), are changed. As is clear from (\ref{m2shift}), the difference between the moments
$M_n(Q^2)$ at the normalization points ${\bar m}^2$ and $Q^2+{\bar m}^2$ increases with $Q^2$.
At $Q^2$ used above, the difference is moderate and, if recalculated to $\left< {\alpha_s\over \pi}G^2\right>$,
results in the error $\Delta \left< {\alpha_s\over \pi}G^2\right>\sim 2\times 10^{-3} \, {\rm GeV}^4$,
much smaller, than the overall error in (\ref{mgrestr}). However, going to the higher $Q^2$ would be dangerous.
In fact, while deriving (\ref{m2shift}), we expanded the running QCD coupling $a(\mu^2)$ in
$a({\bar m}^2)$. This expansion is valid if
\be
a\, \beta_0\ln{Q^2\over {\bar m}^2} \, \ll \, 1
\ee
In particular for $Q^2/(4{\bar m}^2)=3$ the l.h.s.~of this equation is $\sim 0.5$ and the neglected higher
order terms could be significant. For this reason we avoid to use higher $Q^2$, than it was done.

Let us now turn the problem around and try to predict the width $\Gamma_{J/\psi\to ee}$ theoreticaly.
In order to avoid the wrong circle argumentation we do not use the condensate value just obtained,
but take the limitation $\left< {\alpha_s\over \pi}G^2\right>=0.006\pm 0.012 \, {\rm GeV}^4$ found
in \cite{GIZ} from $\tau$-decay data. Then, the mass limits ${\bar m}=1.28 -1.33\,{\rm GeV}$ can be found from the
moment ratios exhibited above, which do not depend on $\Gamma_{J/\psi\to ee}$
if the contributions of higher resonances is approximated by continuum (the accuracy of such approximation
is about $3\%$). The substitution of these values of ${\bar m}$ into the moments gives
\be
\label{jpsiwt}
\Gamma^{\rm theor}_{J/\psi\to ee}\,=\,4.9 \pm 0.8 \, {\rm keV}
\ee
in comparison with experimental value $\Gamma_{J/\psi\to ee}=5.26\pm 0.37\,{\rm keV}$. Such good
coincidence of the theoretical prediction and experimental data is a very impressive demonstration of the
QCD sum rules effectiveness. It must be stressed, that while obtaining (\ref{jpsiwt}) no additional
input were used besides the condensate restriction taken from \cite{GIZ} and the value of $\alpha_s(m_\tau^2)$.

%%%%%%%%%%%%%%%%%%%% D=6 CONDENSATE %%%%%%%%%%%%%%%%%%%%

\section{{\boldmath $D=6$} condensate influence}

The $D=4$ gluon  condensate $\left<aG^2\right>$ is the leading term in the
operator expansion series.  The question arises, how the higher dimension condensate
could change the results of our analysis. There is single $D=6$  gluon condensate
$\left<g^3G^3\right>$. Its contribution to the polarization function (\ref{podef})
can be parametrized as follows:
$$
\Pi^{(G3)}(s)\, = \,{\left< g^3f^{abc}G^a_{\mu\nu}G^b_{\nu\lambda}G^c_{\lambda\mu}\right>
  \over (4m^2)^3}\,f^{(G3)}(z)  \; ,\qquad z={s\over 4m^2} \; ,
$$
The dimensionless function $f^{(G3)}(z)$ has been found in \cite{NR}:
\be
f^{(G3)}(z)\,=\,-\,{1\over 72\,\pi^2 z^3} \,\left( \,{2\over 15}\,+\,{2\over 5}\,z\,
+\,4\,J_2 \,-\,{31\over 3}\,J_3\,+\,{43\over 5}\,J_4\,-\,{12\over 5}\,J_5 \,\right) \; ,
\ee
where the integrals
$$
J_n\,=\,\int_0^1 {dx\over \left[ 1-4zx(1-x)\right]^n}
$$
can be calculated analytically.  However the integral representation is convenient
to express the result in terms of Gauss hypergeometric function, which can be easily
differentiated in order to obtain the moments:
\be
\label{d6gcm}
M_n(Q^2)\,=M_n^{(0)}(Q^2)\,+\,\ldots\,+\,
 \left< g^3f^{abc}G^a_{\mu\nu}G^b_{\nu\lambda}G^c_{\lambda\mu}\right>\,M^{(G3)}_n(Q^2)  \; ,
\ee
where
 $$
 M_n^{(G3)}(Q^2)\,=\, {\sqrt{\pi} \over 1080\,(4m^2)^{n+3}}\, \sum_{i=2}^4
 c_i\,{\Gamma(n+i)\,\Gamma(n+5)\over \Gamma(n+1)\,\Gamma(n+9/2)}
 \,{}_2\!F_1\!\left( \left.{n+i,\,n+5\atop n+9/2}\right| -\,{Q^2\over 4m^2}  \right)
$$
and the constants $c_2=3$, $c_3=-7$, $c_4=-9$. Significance of the condensate $\left<g^3G^3\right>$
is determined by the ratio of the two terms in (\ref{d6gcm}). The numerical values of this ratio
for different $(n,Q^2)$ are given in the last column of the Tables 1,2,3 in the Appendix.

No reliable estimations of the $\left<G^3\right>$ condensate are available. There
exists only the estimation, based on the dilute instanton gas  model  \cite{NSVZ}:
\be
\label{g3inst}
 \left< g^3f^{abc}G^a_{\mu\nu}G^b_{\nu\lambda}G^c_{\lambda\mu}\right>  \,=\,
{4\over 5} \,{12 \pi^2 \over \rho_c^2} \left< {\alpha_s\over\pi} G^2\right> \; ,
\ee
where $\rho_c$ is effective instanton radius. The numerical value of $\rho_c$ is uncertain,
even in the framework of the model: in \cite{SS} the value $\rho_c= 1/3 \,{\rm fm}=1.5\,{\rm GeV}^{-1}$
was advocated, in \cite{SVZ} the value $\rho_c= 1 \,{\rm fm}=4.5\,{\rm GeV}^{-1}$ was used.
In the recent paper \cite{IS}, based on the sensitive to gluon condensate sum rules,
$\rho_c= 0.5 \,{\rm fm}=2.5\,{\rm GeV}^{-1}$ was suggested.

\begin{figure}[tb]
\hspace{41mm} \epsfig{file=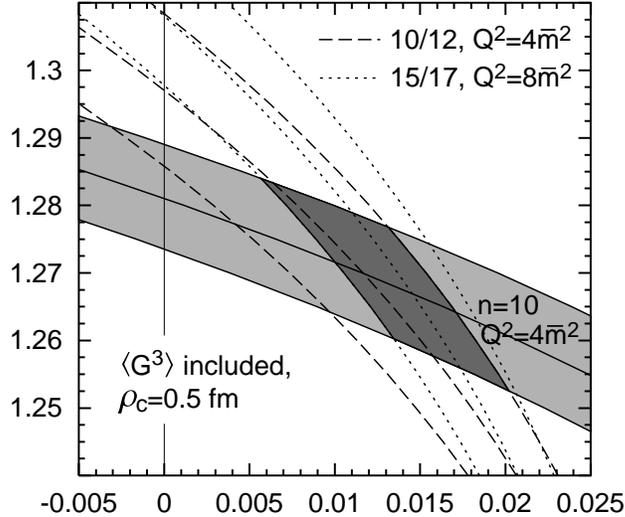, width=82mm}
\caption{$\ov{\rm MS}$ mass versus gluon condensate obtained from
the moments and ratios with account of $\left<G^3\right>$ condensate according to (\ref{g3inst}).}
\label{mcd6}
\end{figure}

The contribution of $\left<g^3G^3\right>$ to $M_n(Q^2)$ at a fixed $n$ falls rapidly with
growth of $Q^2$. At $Q^2=0$ and $n\ge 5$ it comprises about $50\%$ or more of the
gluon condensate contribution at $\rho_c= 0.5 \,{\rm fm}$. Even at $Q^2/(4{\bar m}^2)=1$
it is significant: the (negative) correction to the gluon condensate term is $\sim 10\%$
in $M_{10}$ and $\sim 30\%$ in the ratio $M_{10}/M_{12}$. One gets more reliable
results at   $Q^2/(4{\bar m}^2)=2$. Here the corrections are: $-7\%$ for $M_{15}$ and
$-18\%$ for $M_{15}/M_{17}$. These corrections leave the charm quark mass
almost unchanged, but increase the gluon condensate and its error (compare Figs
\ref{mc12} and \ref{mcd6}). The account of $\left<g^3G^3\right>$ contribution
leads to the following restriction:
\be
\left< {\alpha_s\over \pi} G^2\right>\,=\,0.011\pm 0.009 \, {\rm GeV}^4
\ee
Certainly, it relies upon the instanton gas model, that gives (\ref{g3inst}).

%%%%%%% ABOUT THE ATTEMPTS...%%%%%%%%%%%%%%%%%%%

\section{About the attempts to sum up the Coulomb-like corrections}

Sometimes when considering of the heavy quarkonia sum rules the Coulomb-like corrections
are summed up  \cite{JP}, \cite{CHKST}, \cite{N}--\cite{KSh}.
The basic argumentation for such summation
is that at $Q^2=0$ and high $n$ only small quark velocities $v\lesssim 1/\sqrt{n}$ are
essential and the problem becomes nonrelativistic.
So it is possible to perform the summation with the help of well known formulae
of nonrelativistic quantum mechanics for $|\psi(0)|^2$ in case of Coulomb interaction
(see \cite{LL}).

This method was not used here for the following reasons:

1. The basic idea of our approach is to calculate the moments of the polarization operator
in QCD by applying the perturbation theory and OPE (l.h.s.~of the sum rules) and to
compare it with the r.h.s.~of the sum rules, represented by the contribution of charmonium
states (mainly by $J/\psi$). Therefore it is assumed, that the theoretical side of the sum rule
is dual to experimental one, i.e. the same domains of coordinate and momentum spaces are
of importance at both sides. But the charmonium states (particularly, $J/\psi$) are by no
means the Coulomb systems. A particular argument in favor of this statement is the
ratio $\Gamma_{J/\psi\to ee}/\Gamma_{\psi'\to ee}=2.4$. If charmonia were
nonrelativistic Coulomb system, $\Gamma_{\psi\to ee}$ would be proportional to
$|\psi(0)|^2\sim 1/(n_r+1)^3$, and since $\psi'$ is the first radial excitation with
$n_r=1$, this ratio would be equal to 8 (see also \cite{LL}).

2. The heavy quark-antiquark Coulomb interaction at large distances
$r>r_{\rm conf}\sim 1 \,{\rm GeV}^{-1}$ is screened by gluon and light quark-antiquark clouds,
resulting in string formation. Therefore the summation of Coulombic series makes
sense only when the Coulomb radius $r_{\rm Coul}$ is below $r_{\rm conf}$. (It must be taken
in mind, that higher order terms in Coulombic series represent the contributions of large distances,
$r\gg r_{\rm Coul}$.) For charmonia we have
$$
r_{\rm Coul}\,\approx\, {2\over m_c C_F \alpha_s}\, \approx  \, 4 \,{\rm GeV}^{-1}
$$
It is clear, that the necessary condition $R_{\rm Coul} < R_{\rm conf}$ is badly violated
for charmonia. This means that the summation of the Coulomb series in case of charmonium
would be a wrong step.

3. Our analysis is performed at $Q^2/4{\bar m}^2\ge 1$.
At large $Q^2$ the Coulomb corrections are suppressed in comparison with $Q^2=0$. It is easy to estimate
the characteristic values of the quark velocities. At large $n$ they are $v\approx\sqrt{(1+Q^2/4m^2)/n}$.
We are working along the diagonals of the Fig \ref{msb}, well parametrized by the equation
 $Q^2/4{\bar m}^2 \approx n/5-1$. Here the quark velocity $v\sim 1/\sqrt{5}\approx 0.45$ is not small
 and not in the nonrelativistic domain, where the Coulomb corrections are large and legitimate.

 Nevertheless let us look on the expression of $R_c$, obtained after summation of the Coulomb corrections in the
 nonrelativistic theory \cite{E}. It reads (to go from QED to QCD one has to replace $\alpha\to C_F\alpha_s$,
 $C_F=4/3$):
 \be
 R_{c, \,{\rm Coul}}\,=\,{3\over 2} \, {\pi C_F \alpha_s\over 1 - e^{-x}}  \,=\,
 {3\over 2} \, v \left( \, 1\,+ \,{x\over 2}\,+ \,{x^2\over 12} \,-\,{x^4\over 720}\,+\,\ldots\, \right)
 \label{rcoul}
 \ee
 where $x=\pi C_F\alpha_s/v$. At $v= 0.45$ and $\alpha_s\approx 0.26$ the first 3 terms in the expansion
 (\ref{rcoul}), accounted in our calculations, reproduce the exact value of $R_{c,\, {\rm Coul}}$ with accuracy
 $1.6\%$. Such deviation leads to the error of the mass ${\bar m}$ of order $(1-2)\times 10^{-3}\,{\rm GeV}$,
 which is completely negligible. In order to avoid misunderstanding, it must be mentioned, that the value of
 $R_{c\, {\rm Coul}}$, computed by summing the Coulomb corrections in nonrelativistic theory has not too
 much in common with real physical situation. Numerically, at choosen values of the parameters,
 $R_{c\, {\rm Coul}}\approx 1.8$, while the real value (both experimantal and in the perturbative QCD)
 is about $1.1$. The goal of the argumens, presented above, was to demonstrate, that even in the case of
 Coulombic system  our approach would have a good accuracy of calculation.

At $v=0.45$ the momentum transfer from quark to antiquark
 is $\Delta p \sim 1\,{\rm GeV}$. (This is typical domain for
QCD sum rule validity.) In coordinate space it corresponds to
$\Delta r_{q {\bar q}} \sim 1\, {\rm GeV}^{-1}$. Comparison with potential models
\cite{E} demonstrates, that in this region the effective potential strongly differs from
Coulombic one.

4. Large compensation of various terms in the expression for the moments in $\ov{\rm MS}$
scheme (see Fig \ref{fig_mdm}) is not achieved, if only the Coulomb terms are taken into
account. This means, that the terms of non-Coulombic origin are more important here, than
Coulombic ones.

For all these reasons we believe, that the summation of nonrelativistic Coulomb
corrections is inadequate in the problem in view: it will not improve the accuracy of
calculations, but would be misleading.

%%%%%%%%%%%%%%%% RESULTS... %%%%%%%%%%%%%%%%%%

\section{Results and discussion}

The analysis of charmonium sum rules is performed within the framework of QCD at the next level
of precission in comparison with famous treatment of this problem by Shifman, Vainstein and
Zakharov \cite{SVZ}. In the perturbation theory the terms of order $\alpha_s^2$ were accounted as well
as $\alpha_s$ corrections to the gluon condensate contribution,
in OPE --- the dimension 6 operator $G^3$. The method of the moments was exploited.
The validity of the method  was  demonstrated for the $\ov{\rm MS}$ mass of the charm quark,
but not for the pole mass. The domain in the plane $(n,Q^2)$ was found, where the three accounted terms
in the perturbative series are well converging. It was shown, that the sum rules do not work at $Q^2=0$,
where the following 4 requirements cannot be satisfied simultaneously: 1) convergence of the perturbation
series, 2) small $\alpha_s$ correction to the gluon condensate contribution, 3)
small contribution of $G^3$ operator, 4) small contribution of higher resonances and continuum.
Large $Q^2$ allow also to suppress the Coulomb corrections.
The most suitable values of $Q^2$ for the sum rules are
$Q^2 \sim (1-2) 4{\bar m}^2 \sim 5-15 \,{\rm GeV}^2$. The values of charmed quark $\ov{\rm MS}$ and the
gluon condensate were found by comparing the theoretical moments with experimental ones,
saturated by charmonium resonances (plus continuum). A strong correlation of the values
${\bar m}$ and $\left< {\alpha_s\over \pi} G^2\right>$ was established. This connection only weakly
depends on $\alpha_s$. Taking the $\alpha_s$ value found in \cite{GIZ} from
hadronic $\tau$-decay data
\be
\label{alphatau2}
\alpha_s(m_\tau^2)\,=\,0.330\pm 0.025 \, ,
\ee
 the $\ov{\rm MS}$ charm quark mass and the gluon condesate were determined
\be
\label{mgrestr2}
{\bar m}({\bar m}^2)\,=\,1.275\pm 0.015 \, {\rm GeV} \; , \qquad
\left< {\alpha_s\over \pi} G^2\right>\,=\,0.009\pm 0.007 \, {\rm GeV}^4
\ee
The error in (\ref{mgrestr2}) roughly comprises as $50\%$ theoretical (uncertainly in $\alpha_s$
and the normalization scale) and $50\%$ experimental (mainly the error of $J/\psi$ electronic
decay width). The numbers in (\ref{mgrestr2}) were obtained disregarding the contribution of $G^3$
operator. The account of $G^3$ term, when $\left< G^3 \right>$ was taken using the dilute
instanton gas model with $\rho_c=0.5\,{\rm fm}$, shifts (\ref{mgrestr2}) to
\be
\label{grestr1}
\left< {\alpha_s\over \pi} G^2\right>\,=\,0.011\pm 0.009 \, {\rm GeV}^4
\ee
The value (\ref{grestr1}) may be compared with recently found \cite{GIZ} limitation on the
gluon condensate from hadronic $\tau$-decay data:
\be
\label{grestr2}
\left< {\alpha_s\over \pi} G^2\right>\,=\,0.006\pm 0.012 \, {\rm GeV}^4
\ee
Eqs (\ref{grestr1}) and (\ref{grestr2}) are compatible and obtained from independent sources.
So, with some courage, we can average them and get
\be
\label{grestr3}
\left< {\alpha_s\over \pi} G^2\right>_{\rm av}\,=\,0.0085\pm 0.0075 \, {\rm GeV}^4
\ee
After such averaging we come back to (\ref{mgrestr2}).

We can formulate our final conclusion about the gluon condensate value in such a way.
The values of gluon condensate two times (or more) larger than the SVZ value (\ref{svzval}) are
certainly excluded. Unfortunately our analysis does not allow to exclude zero values of the gluon condensate.
In this aspect the improvement of the experimental precission of  $J/\psi\to e^+e^-$ width
would be helpfull. Based on the condensate limitation (\ref{grestr2}) and the value of $\alpha_s$
(\ref{alphatau2}), the $J/\psi$ electronic decay width $\Gamma_{J/\psi\to ee}$ was predicted theoretically:
\be
\label{jpsiwt2}
\Gamma^{\rm theor}_{J/\psi\to ee}\,=\,4.9 \pm 0.8 \, {\rm keV}
\ee
in comparison with the experimental value $5.26\pm 0.37 \,{\rm keV}$. Such a good coincidense ones
more demonstrates the effectiveness of QCD sum rule approach.

\section*{Acknowledgement}

Authors thank A.I. Vainstein and K.G. Chetyrkin for fruitful discussions.

The research described in this publication was made possible in part by Award No RP2-2247
of U.S. Civilian Research and Development Foundation for Independent States of Former
Soviet Union (CRDF), by the Russian Found of Basic Research grant 00-02-17808 and
INTAS grant 2000, project 587.

\section*{Appendix: Numerical values of the moments}

We list here the numerical values of the perturbative moments ${\bar M}^{(0,1,2)}$,
condensate contribution ${\bar M}^{(G,0,1)}$ in $\ov{\rm MS}$ scheme
computed by (\ref{mommsb}) and $\left<G^3\right>$ condensate contribution
$M^{(G3)}$ (\ref{d6gcm})  for $n=1..20$ and $Q^2/(4{\bar m}^2)=0,1,2$.
For dimensionfull values we put $4{\bar m}^2=1$ here, so that
the leading term ${\bar M}^{(0)}$ and the ratios ${\bar M}^{(G,0)}/{\bar M}^{(0)}$,
${\bar M}^{(G3)}/{\bar M}^{(0)}$ should be divided by
 $(4{\bar m}^2)^n$ and $(4{\bar m}^2)^2$, $(4{\bar m}^2)^3$ respectively for a
particular mass ${\bar m}$.

\begin{table}[ht]
$$
\ba{|c|ccc|cc|c|} \hline
n & {\bar M}_n^{(0)} & {\bar M}_n^{(1)}\!/{\bar M}_n^{(0)} & {\bar M}_n^{(2)}\!/{\bar M}_n^{(0)}
  & {\bar M}_n^{(G,0)}\!/{\bar M}_n^{(0)}\! & \!{\bar M}_n^{(G,1)}\!/{\bar M}_n^{(G,0)} &
 {\bar M}_n^{(G3)}\!/{\bar M}_n^{(0)}  \\ \hline
1 &0.8 & 2.394 & 2.384 & -15.04 & 2.477 & 0.056\\
2 &0.3429 & 2.427 & 6.11 & -58.49 & 1.054 & 0.826\\
3 &0.2032 & 1.917 & 6.115 & -143.6 & -0.484 & 4.003\\
4 &0.1385 & 1.1 & 4.402 & -283.4 & -2.107 & 12.76\\
5 &0.1023 & 0.078 & 2.162 & -491.3 & -3.798 & 32.21\\
6 &7.9565\times 10^{-2} & -1.092 & 0.213 & -780.3 & -5.545 & 69.81\\
7 &6.4187\times 10^{-2} & -2.375 & -0.836 & -1164. & -7.337 & 135.9\\
8 &5.3207\times 10^{-2} & -3.75 & -0.514 & -1654. & -9.17 & 243.9\\
9 &4.5043\times 10^{-2} & -5.199 & 1.559 & -2266. & -11.04 & 411.2\\
10 &3.8776\times 10^{-2} & -6.711 & 5.698 & -3011. & -12.94 & 659.1\\
11 &3.3841\times 10^{-2} & -8.277 & 12.17 & -3903. & -14.86 & 1014.\\
12 &2.9872\times 10^{-2} & -9.89 & 21.19 & -4955. & -16.81 & 1506.\\
13 &2.6624\times 10^{-2} & -11.54 & 32.98 & -6181. & -18.78 & 2172.\\
14 &2.3924\times 10^{-2} & -13.23 & 47.69 & -7593. & -20.77 & 3055.\\
15 &2.1653\times 10^{-2} & -14.96 & 65.49 & -9204. & -22.78 & 4204.\\
16 &1.9719\times 10^{-2} & -16.71 & 86.51 & -1.103\times 10^{4} & -24.81 & 5673.\\
17 &1.8058\times 10^{-2} & -18.49 & 110.9 & -1.308\times 10^{4} & -26.85 & 7526.\\
18 &1.6617\times 10^{-2} & -20.3 & 138.7 & -1.537\times 10^{4} & -28.91 & 9834.\\
19 &1.5359\times 10^{-2} & -22.13 & 170.1 & -1.791\times 10^{4} & -30.98 & 1.268\times 10^{4}\\
20 &1.4252\times 10^{-2} & -23.98 & 205.2 & -2.072\times 10^{4} & -33.07 & 1.614\times 10^{4}\\
 \hline
\ea
$$
\caption{Moments at $Q^2=0$}
\end{table}

\begin{table}
$$
\ba{|c|ccc|cc|c|} \hline
n & {\bar M}_n^{(0)} & {\bar M}_n^{(1)}\!/{\bar M}_n^{(0)} & {\bar M}_n^{(2)}\!/{\bar M}_n^{(0)}
  & {\bar M}_n^{(G,0)}\!/{\bar M}_n^{(0)}\! & \!{\bar M}_n^{(G,1)}\!/{\bar M}_n^{(G,0)} &
 {\bar M}_n^{(G3)}\!/{\bar M}_n^{(0)}  \\ \hline
1 &0.4348 & 2.235 & -0.307 & -2.816 & 4.532 & -0.02\\
2 &9.7902\times 10^{-2} & 2.64 & 4.407 & -10.19 & 4.03 & -0.058\\
3 &2.9985\times 10^{-2} & 2.709 & 6.752 & -23.8 & 3.455 & -0.082\\
4 &1.047\times 10^{-2} & 2.588 & 7.653 & -45.32 & 2.825 & -0.014\\
5 &3.9365\times 10^{-3} & 2.34 & 7.582 & -76.4 & 2.15 & 0.279\\
6 &1.5529\times 10^{-3} & 1.999 & 6.85 & -118.7 & 1.438 & 1.004\\
7 &6.3364\times 10^{-4} & 1.587 & 5.683 & -173.9 & 0.697 & 2.452\\
8 &2.6515\times 10^{-4} & 1.118 & 4.253 & -243.5 & -0.071 & 5.015\\
9 &1.1314\times 10^{-4} & 0.601 & 2.7 & -329.4 & -0.862 & 9.197\\
10 &4.9032\times 10^{-5} & 0.045 & 1.136 & -433. & -1.673 & 15.63\\
11 &2.1523\times 10^{-5} & -0.546 & -0.343 & -556.1 & -2.501 & 25.09\\
12 &9.5483\times 10^{-6} & -1.167 & -1.656 & -700.3 & -3.346 & 38.5\\
13 &4.2743\times 10^{-6} & -1.815 & -2.732 & -867.3 & -4.205 & 56.96\\
14 &1.9283\times 10^{-6} & -2.486 & -3.508 & -1059. & -5.078 & 81.75\\
15 &8.7574\times 10^{-7} & -3.178 & -3.93 & -1276. & -5.962 & 114.4\\
16 &4.0007\times 10^{-7} & -3.89 & -3.948 & -1521. & -6.858 & 156.4\\
17 &1.8372\times 10^{-7} & -4.62 & -3.518 & -1795. & -7.764 & 209.9\\
18 &8.4756\times 10^{-8} & -5.365 & -2.599 & -2101. & -8.68 & 277.\\
19 &3.9264\times 10^{-8} & -6.126 & -1.154 & -2439. & -9.605 & 360.\\
20 &1.8257\times 10^{-8} & -6.9 & 0.85 & -2811. & -10.54 & 461.7\\
\hline
\ea
$$
\caption{Moments at $Q^2=4 {\bar m}^2$}
\end{table}

\begin{table}
$$
\ba{|c|ccc|cc|c|} \hline
n & {\bar M}_n^{(0)} & {\bar M}_n^{(1)}\!/{\bar M}_n^{(0)} & {\bar M}_n^{(2)}\!/{\bar M}_n^{(0)}
  & {\bar M}_n^{(G,0)}\!/{\bar M}_n^{(0)}\! & \!{\bar M}_n^{(G,1)}\!/{\bar M}_n^{(G,0)} &
 {\bar M}_n^{(G3)}\!/{\bar M}_n^{(0)}  \\ \hline
1 &0.3005 & 2.073 & -2.016 & -1.098 & 5.002 & -8.936\times 10^{-3}\\
2 &4.6172\times 10^{-2} & 2.531 & 2.453 & -3.825 & 4.762 & -0.03\\
3 &9.589\times 10^{-3} & 2.734 & 5.208 & -8.691 & 4.468 & -0.064\\
4 &2.2613\times 10^{-3} & 2.792 & 6.909 & -16.2 & 4.131 & -0.105\\
5 &5.7274\times 10^{-4} & 2.753 & 7.853 & -26.84 & 3.761 & -0.14\\
6 &1.5191\times 10^{-4} & 2.643 & 8.229 & -41.11 & 3.364 & -0.147\\
7 &4.1621\times 10^{-5} & 2.478 & 8.167 & -59.5 & 2.942 & -0.094\\
8 &1.1682\times 10^{-5} & 2.27 & 7.769 & -82.51 & 2.501 & 0.066\\
9 &3.3404\times 10^{-6} & 2.025 & 7.113 & -110.6 & 2.042 & 0.393\\
10 &9.6957\times 10^{-7} & 1.749 & 6.264 & -144.3 & 1.568 & 0.963\\
11 &2.8488\times 10^{-7} & 1.447 & 5.276 & -184.1 & 1.08 & 1.871\\
12 &8.4559\times 10^{-8} & 1.122 & 4.195 & -230.4 & 0.579 & 3.233\\
13 &2.5317\times 10^{-8} & 0.776 & 3.061 & -283.9 & 0.068 & 5.186\\
14 &7.6361\times 10^{-9} & 0.412 & 1.909 & -344.8 & -0.455 & 7.89\\
15 &2.3181\times 10^{-9} & 0.031 & 0.77 & -413.8 & -0.986 & 11.53\\
16 &7.0768\times 10^{-10} & -0.364 & -0.33 & -491.3 & -1.527 & 16.33\\
17 &2.1713\times 10^{-10} & -0.773 & -1.365 & -577.8 & -2.075 & 22.52\\
18 &6.6914\times 10^{-11} & -1.195 & -2.313 & -673.9 & -2.631 & 30.38\\
19 &2.0704\times 10^{-11} & -1.628 & -3.153 & -779.9 & -3.194 & 40.21\\
20 &6.429\times 10^{-12} & -2.072 & -3.867 & -896.4 & -3.764 & 52.36\\
\hline
\ea
$$
\caption{Moments at $Q^2=8 {\bar m}^2$}
\end{table}

\end{document}